\documentstyle[preprint,tighten,aps]{revtex}

\begin{document}
\draft

\title{Bell's inequality tests: from photons to $B$--mesons}

\author{A.~Bramon$^1$, R.~Escribano$^{1,2}$ and G.~Garbarino$^3$}

\address{$^1$Grup de F{\'\i}sica Te\`orica, Universitat Aut\`onoma de Barcelona, 
E-08193 Bellaterra, Barcelona, Spain} 

\address{$^2$IFAE, Universitat Aut\`onoma de Barcelona,
E-08193 Bellaterra, Barcelona, Spain}

\address{$^3$Dipartimento di Fisica Teorica, Universit\`a di
Torino and INFN, Sezione di Torino, I-10125 Torino, Italy}

\date{\today}
\maketitle

\begin{abstract}
We analyse the recent claim that a violation of a Bell's inequality has been observed 
in the $B$--meson system [A. Go, {\em Journal of Modern Optics} {\bf 51} (2004) 991]. 
The results of this experiment are a convincing proof of quantum entanglement in
$B$--meson pairs similar to that shown by polarization entangled photon pairs. 
However, we conclude that the tested inequality is not a genuine
Bell's inequality and thus cannot discriminate between quantum mechanics and local
realistic approaches.  
\end{abstract}


\newpage
\pagestyle{plain}
\baselineskip 16pt
\vskip 48pt

\newpage

\date{\today}

A recent paper \cite{Go} claims that a clear violation of a Bell's
inequality \cite{Bell,CHSH} has been observed using particle--antiparticle correlations
in semileptonic $B$--meson decays. The $B$--meson pairs were produced via 
$\Upsilon(4S)\rightarrow B^0\bar B^0$ decays at Belle with a wave function which is
formally identical to that of polarization entangled photon pairs. The experiment
\cite{Go} is the first attempt to perform a Bell's inequality test with high--energy
particles ---rather than with the more conventional photons in optical tests--- and thus
deserves special attention \cite{PW}. The purpose of this Letter is to show that the
results of Ref.~\cite{Go}, which conclusively prove the entangled behaviour of such
heavy--meson pairs, cannot be considered a genuine Bell--test, i.e.~a test
discriminating between local realism (LR) and quantum mechanics (QM).

Just after the $\Upsilon(4S)$ decay ($t=0$), the state of the $B^0\bar B^0$ pair is
\begin{equation}
\label{1} |\Upsilon(0)\rangle  =  \frac{1}{\sqrt 2}\left[
|B^0\rangle_l |\bar{B}^0\rangle_r - |\bar{B}^0\rangle_l
|B^0\rangle_r\right]
 =  \frac{1}{\sqrt 2}\left[
|B_L\rangle_l |B_H\rangle_r - |B_H\rangle_l |B_L\rangle_r\right]\ ,
\end{equation}
where $l$ and $r$ denote the `left' and `right' directions of
motion of the two separating mesons and (small) $CP$--violating effects have been  
(safely \cite{PDG}) neglected in the last equality.
This approximation implies $|B_L\rangle=\{|B^0\rangle+|\bar B^0\rangle\}/\sqrt{2}$ 
and $|B_H\rangle=\{|B^0\rangle-|\bar B^0\rangle\}/\sqrt{2}$. 
Note the strong similarity between the state (\ref{1}) and
the singlet state used in two--photon polarization Bell--tests. According to QM, the time
evolution in free space of the light-- and heavy--mass eigenstates, $B_L$ and $B_H$, is 
given by
\begin{equation}
\label{2} 
|B_{L,H}\rangle \rightarrow e^{-i m_{L,H} t} 
e^{-\frac{1}{2}\Gamma_{L,H} t} |B_{L,H}\rangle\ ,
\end{equation} 
where $\Gamma_{L,H}$ account for the common $B_L$ and $B_H$ decay rates. 
Experimentally one has $\Gamma_{L} \simeq \Gamma_{H} = 1 /\tau_B$ 
and $\tau_B=(1.536\pm 0.014)\times 10^{-12}$ s \cite{Go,PDG}.
The evolution of the initial state (\ref{1}) up to left-- and right--side times $t_l $
and $ t_r$ is then
\begin{eqnarray}
\label{3} 
|\Upsilon(t_l ; t_r)\rangle &=& \frac{1}{2\sqrt{2}}
e^{-\frac{t_l +t_r}{2\tau_B}} 
\left\{ (1-e^{i \Delta m (t_l -t_r)} )
\left[|B^0\rangle_l|B^0\rangle_r-|\bar B^0\rangle_l|\bar
B^0\rangle_r \right] \right. \nonumber \\  
&& +\left. (1+e^{i \Delta m (t_l -t_r)} ) \left[|B^0\rangle_l|\bar
B^0\rangle_r-|\bar B^0\rangle_l|B^0\rangle_r\right] \right\}\ , 
\end{eqnarray}
where $\Delta m\equiv m_H-m_L\ne 0$ induces flavour or $B^0$--$\bar B^0$ oscillations in
time. These oscillations are crucial for a Bell--test and, formally, they play the same 
role as the different orientations of polarization analysers in photonic experiments 
(see Refs.~\cite{GoGisin,BN}). Eq.~(\ref{3}) leads to the same-- and opposite--flavour joint
detection probabilities 
\begin{eqnarray}
\label{4}
P_{B^0 B^0}(t_l ; t_r) =  P_{\bar B^0 \bar B^0}(t_l ; t_r) = 
\frac{1}{4} e^{-\frac{t_l +t_r}{\tau_B}} \left[ 1- \cos (\Delta m \, \Delta t)\right]\ , &&
\nonumber \\
P_{B^0 \bar B^0}(t_l ; t_r) =  P_{\bar B^0 B^0}(t_l ; t_r) = 
\frac{1}{4} e^{-\frac{t_l +t_r}{\tau_B}} \left[ 1+ \cos (\Delta m \, \Delta t)\right]\ , &&  
\end{eqnarray}
with $\Delta t \equiv t_l -t_r$. 
These $P_{B_l B_r}(t_l ; t_r)$'s, with $B_{l,r}=B^0, \bar B^0$, are the QM predictions
for the joint probability  measurements performed by two \emph{hypothetical} flavour 
detectors inserted along the left and right beams at `time--of--fight distances' $t_l$
and $t_r$. 

But no such detectors are available and the flavour (either $B^0$ or $\bar B^0$)
of each member of the $B$--meson pair has to be identified by observing its
decay modes. The various decay products $f=  D^*(2010)^- l^+ \nu_l$, $D^- \pi^+,\dots$, 
which are forbidden for a $\bar B^0$, unambiguously come from a
$B^0$, while the opposite is true for the respective charge conjugated modes 
$\bar f=  D^*(2010)^+ l^- \bar \nu_l$, $D^+ \pi^-,\dots$ ($l^\pm= e^\pm, \mu^\pm$).
The corresponding partial decay widths satisfy 
$\Gamma_{B^0 \to f} = \Gamma_{\bar B^0 \to \bar f}$ \cite{PDG}.
Experimentally, one counts the number of joint $B$--meson decay events
into the distinct decay modes $f_{l,r}$ and 
in the appropriate time intervals $[t_{l,r}, t_{l,r} + dt_{l,r}]$; then the
joint decay probabilities ${\cal P}_{f_l,f_r}(t_l ; t_r)$ are obtained after 
dividing these numbers by the total number of initial $B^0\bar B^0$ pairs. 
Finally, the corresponding joint decay rates $\Gamma_{f_l,f_r}(t_l ; t_r)$ are derived as:
\begin{eqnarray}
\label{5}
\Gamma_{f_l,f_r}(t_l ; t_r)\equiv
\frac{d^2 {\cal P}_{f_l,f_r}(t_l ; t_r)}{dt_l\; dt_r} = P_{B_l B_r}(t_l ; t_r)\;
\Gamma_{B_l \to f_l}\; \Gamma_{B_r \to f_r}\ ,
\end{eqnarray}
from which the joint probabilities $P_{B_l B_r}(t_l ; t_r)$ immediately
follow. The data from Ref.~\cite{Go}, where the only detected modes were 
$B^0 \to D^*(2010)^- l^+ \nu_l$ and $\bar B^0 \to  D^*(2010)^+ l^- \bar\nu_l$ occurring
with a 5.4\% probability each, are found to be in good agreement with the QM predictions 
in Eq.~(\ref{4}). This is a convincing proof of the entanglement between the members of each 
$B$--meson pair and thus suggests the possibility of interesting Bell--tests.

To this end, one proceeds as in optical experiments and defines the normalized
correlation function \cite{Go} 
\begin{equation}
\label{6}
 E_R(\Delta t ) \equiv
\frac{P_{B^0 B^0} (\Delta t) + P_{\bar B^0 \bar B^0} (\Delta t) - 
 P_{B^0 \bar B^0} (\Delta t) - P_{\bar B^0 B^0} (\Delta t)} 
 {P_{B^0 B^0} (\Delta t) + P_{\bar B^0 \bar B^0} (\Delta t) + 
 P_{B^0 \bar B^0} (\Delta t) + P_{\bar B^0 B^0} (\Delta t)} \\
= -\cos (\Delta m \, \Delta t)\ ,
\end{equation}
which turns out to depend on $\Delta t \equiv t_l-t_r$ but not on the specific decay
modes detected in a given experiment. 
In conventional Bell--tests one then derives from LR an inequality which 
has to be necessarily satisfied by these correlation functions. 
Ref.~\cite{Go} makes use of the Clauser, Horne, Shimony and Holt (CHSH) version of the 
Bell inequality \cite{CHSH}
\begin{eqnarray}
\label{8}
S(\Delta t ) = \left|3 E_R(\Delta t ) - E_R(3 \Delta t )\right| \le 2\ ,
\end{eqnarray} 
which turns out to be clearly violated by the data below $\Delta t \simeq 1.7\, \tau_B
\simeq 2.62\, {\rm ps}$.  The maximal violation [$S_{\rm max} = 2.73 \pm 0.19$, occurring 
when $\Delta t = (2 \pm 0.5)\, {\rm ps}$] is more than three $\sigma$'s above the limit in   
Eq.~(\ref{8}) and compatible with the maximal violation predicted by QM 
[$S_{\rm max}^{\rm QM}= 2 \sqrt 2$]. The existence of QM `non--local' correlations is
certainly proved, but, is this really a proof against LR theories 
as claimed in Ref.~\cite{Go}?

The conventional and most convincing procedure to disclaim such a conclusion consists in
constructing a local model of hidden variables which agrees with the quantum mechanical
predictions and thus with the experimental data of Ref.~\cite{Go}. In the present
case, this is easily achieved by simply adapting an original argument introduced by
Kasday \cite{Kasday} in another context. Each $B^0
\bar B^0$ pair is assumed to be produced  at $t=0$ with a set of hidden variables 
$\{t_l, f_l, t_r, f_r\}$ deterministically specifying \emph{ab ovo} 
the future decay times and decay modes of its two members.
Different $B$--meson pairs are indeed supposed to be produced with a probability distribution
coinciding precisely with the joint decay probability
${\cal P}_{f_l,f_r}(t_l ; t_r)$ entering Eq.~(\ref{5}). Note that the conventional
normalization in the hidden variable space, 
$\int d\lambda\, \rho (\lambda )=1$, is now similarly given by 
\begin{eqnarray}
\label{7}
\Sigma_{f_l,f_r} \int dt_l \int dt_r\, \Gamma_{f_l,f_r}(t_l ; t_r) = 1\ ,
\end{eqnarray}
where the time integrals extend from 0 to $\infty$ and the sum to all $B^0$ and $\bar B^0$
decay modes   (Ref.~\cite{Go} considered only a single decay mode, which was
consistently normalized according to Eq.~(\ref{6})). 
Note also that our proposed hidden variable distribution 
function ${\cal P}_{f_l,f_r}(t_l ; t_r)$ reproduces the successful QM description of all the 
measurements in  Ref.~\cite{Go}. 
More importantly, our \emph{ad hoc} LR model also violates
the inequality (\ref{8}) measured there. This proves that the inequality tested
in Ref.~\cite{Go} is not a genuine Bell--inequality, which, by definition, has to be
satisfied in any LR approach. A similar criticism applies to the inequalities derived
in Ref.~\cite{BF} for entangled $K^0\bar K^0$ pairs.

Experiments performed with freely propagating meson--antimeson pairs, whose flavour is 
measured through the identification of their spontaneous decay modes, are not
suitable for Bell--tests. The main reason is that they can be run in a single
experimental setup, not involving alternative settings \cite{BBGH}. Such experiments
then admit a successful QM description of the meson decay modes and decay times which
can be directly used as an equally successful probability distribution of hidden
variables in a LR approach.  In other words, when deriving a Bell's inequality from LR, 
one has to argue that measurement outcomes on
one side cannot be affected by the experimental setting chosen to measure on the other
side. Such counterfactual considerations \cite{CHSH,Redhead} do not apply to
experiments like that of Ref.~\cite{Go} and the tested inequality is useless in order
to discriminate between QM and LR. For this important purpose, other experiments
involving alternative measurement settings are required. Some genuine Bell's inequality
tests for the $K^0 \bar K^0$ system, exploiting the neutral--kaon regeneration effects,
have been proposed by various authors \cite{BN,regenerated}, but their applicability to
the extremely short--lived $B$--mesons seems to offer serious difficulties. 

\section*{Acknowledgements}
This work has been partly supported by EURIDICE HPRN-CT-2002-00311, 
MURST 2001024324\_007, INFN and BFM-2002-02588.


\begin{thebibliography}{100}

\bibitem{Go}
Go, A., 2004, {\em Journal of Modern Optics} {\bf 51}, 991; e-Print Archive:
{\bf quant-ph/0310192}.

\bibitem{Bell}
Bell, J.S., 1964, {\em Physics} {\bf 1}, 195.

\bibitem{CHSH}
Clauser, J.F., Horne, M.A., Shimony, A., and Holt, R.A., 1969,
{\em Phys. Rev. Lett.} {\bf 23}, 880; 
Clauser, J.F., and Shimony, A., 1978, {\em Rep. Prog. Phys.} {\bf 41}, 1881.

\bibitem{PW}
See, for instance, P.~Rodgers, Editor of {\em Physics World}, in
http://physicsweb.org/article/news/7/11/3.  
The book: Afriat, A., and Selleri, F., 1999,
{\em The Einstein, Podolsky and Rosen Paradox in Atomic, Nuclear and Particle Physics}
(New York: Plenum Press), is a complete introduction to the subject. 

\bibitem{PDG}
The Review of Particle Physics,
Eidelman, S., et al., 2004, {\em Phys. Lett.} {\bf B 592}, 1. 

\bibitem{GoGisin}
Gisin, N., and Go, A., 2001, {\em Am. J. Phys.} {\bf 69}, 264.

\bibitem{BN}
Bramon, A., and Nowakowski, M., 1999, {\em Phys. Rev. Lett.} {\bf 83}, 1.

\bibitem{Kasday}
Kasday, L., 1971, Experimental test of quantum predictions for widely separated 
photons. In {\em Foundations of Quantum Mechanics}, 
B. d'Espagnat ed. (New York: Academic Press), p.195. Proceedings of the
International School of Physics `Enrico Fermi', Course IL.

\bibitem{BF}
Benatti, F., and Floreanini, R., 1998, {\em Phys. Rev.} {\bf D 57}, R1332;
Benatti, F., and Floreanini, R., 2000, {\em Eur. Phys. J.} {\bf C 13}, 267.

\bibitem{BBGH}
This is more generally discussed in 
Bertlmann, R.A., Bramon, A., Garbarino, G., and Hiesmayr, B.C., 2004, e-Print Archive:
{\bf quant-ph/0409051} and {\em Phys. Lett.} {\bf A} (in press), 
where an additional argument concerning the non-unitarity of
the $B$-meson time evolution is similarly developed. 

\bibitem{Redhead}
Redhead, M., 1990, {\em Incompleteness, nonlocality and realism}
(Oxford: Oxford University Press).

\bibitem{regenerated}
Eberhard, P.H., 1993, {\em Nucl. Phys.} {\bf B 398}, 155; 
Bramon, A., and Garbarino, G., 2002, {\em Phys. Rev. Lett.} {\bf 88}, 040403;
Bramon, A., and Garbarino, G., 2002, {\em Phys. Rev. Lett.} {\bf 89}, 160401.

\end{thebibliography}
\end{document}